\documentclass[amsmath,superscriptaddress,reprint]{revtex4-2}
\pdfoutput=1
\usepackage{graphicx}
\usepackage{hyperref}

\begin{document}

\title{Quantum emission assisted by energy landscape modification\\ in pentacene-decorated carbon nanotubes}

\author{Zhen~Li}
\email[Corresponding author. ]{zhen.li@riken.jp}
\affiliation{Quantum Optoelectronics Research Team, RIKEN Center for Advanced Photonics, Saitama 351-0198, Japan}
\affiliation{Nanoscale Quantum Photonics Laboratory, RIKEN Cluster for Pioneering Research, Saitama 351-0198, Japan}
\author{Keigo~Otsuka}
\affiliation{Nanoscale Quantum Photonics Laboratory, RIKEN Cluster for Pioneering Research, Saitama 351-0198, Japan}
\author{Daiki~Yamashita}
\affiliation{Quantum Optoelectronics Research Team, RIKEN Center for Advanced Photonics, Saitama 351-0198, Japan}
\author{Daichi~Kozawa}
\affiliation{Quantum Optoelectronics Research Team, RIKEN Center for Advanced Photonics, Saitama 351-0198, Japan}
\author{Yuichiro~K.~Kato}
\email[Corresponding author. ]{yuichiro.kato@riken.jp}
\affiliation{Quantum Optoelectronics Research Team, RIKEN Center for Advanced Photonics, Saitama 351-0198, Japan}
\affiliation{Nanoscale Quantum Photonics Laboratory, RIKEN Cluster for Pioneering Research, Saitama 351-0198, Japan}

\begin{abstract}
Photoluminescent carbon nanotubes are expected to become versatile room-temperature single-photon sources that have applications in quantum information processing. Quantum emission from carbon nanotubes is often induced by localization of excitons or exciton-exciton annihilation. Here, we modify the local energy landscape of excitons by decorating nanoscale pentacene particles onto air-suspended single-walled carbon nanotubes. Directional exciton diffusion from the undecorated region to the decorated site is demonstrated, suggesting exciton trapping induced by local dielectric screening from pentacene particles. Photoluminescence and photon correlation measurements on a representative carbon nanotube reveal enhanced exciton-exciton annihilation and single-photon emission at room temperature. Pentacene particles are shown to promote strong photon antibunching at the decorated site, indicating that noncovalent functionalization using molecules can be an effective approach for energy landscape modification and quantum emission in carbon nanotubes.
\end{abstract}

\maketitle

Single-photon sources are a crucial component for quantum information processing and quantum key distribution \cite{OBrien:2009,Takemoto:2010,Aharonovich:2016,Senellart:2017,He:2018}, and relentless research efforts have been made to pursue the ideal material platform \cite{Kimble:1977,Diedrich:1987,Basche:1992,Lounis:2000,Michler:2000nature,Kurtsiefer:2000,Laurent:2005,Gazzano:2013,Benyoucef:2013,Choi:2014,Castelletto:2014,He:2015,Park:2015an,Tran:2016,Zhou:2018,Wang:2018,Liu:2018}. In recent years, semiconducting single-walled carbon nanotubes (CNTs) have emerged as a promising candidate for single-photon sources because of their stable excitonic states at room temperature \cite{Wang:2005,Maultzsch:2005} and photoluminescence (PL) wavelengths in the telecommunication range \cite{Weisman:2003,Ishii:2015}. Due to limited screening of the Coulomb interaction in the unique one-dimensional structure, tightly bound electron-hole pairs form excitons that exhibit fast diffusion along the CNT axis \cite{Ogawa:1991,Ando:1997,Cognet:2007,Luer:2009,Moritsubo:2010,Ishii:2015,Ma:2015prl}. In most cases, exciton diffusion in a typical CNT leads to spreading of the exciton population along the CNT axis that gives rise to multiphoton emission, while nonradiative recombination predominantly occurs at the quenching sites between the CNT and the substrate \cite{Moritsubo:2010,Xie:2012}.

To enhance photon antibunching and generate single photons from the excitons in CNTs, proper control of the exciton diffusion is necessary. One commonly adopted approach is the exciton localization. Free excitons in covalently functionalized CNTs can be captured at defect-induced deep trapping states, where PL brightening \cite{Ghosh:2010Science,Miyauchi:2013,Piao:2013,Ma:2014}, high-purity single-photon emission and long exciton lifetimes have been achieved \cite{Ma:2015NatNano,He:2017,Ishii:2018}. Furthermore, spontaneous localization of the band-edge excitons due to environmental disorders or unintentional molecular adsorption is also possible in undoped CNTs at cryogenic temperatures, where signatures of quantum correlation have been observed \cite{Htoon:2004,Hirori:2006,Hogele:2008,Newman:2012,Hofmann:2013,Sarpkaya:2013,Vialla:2014,Endo:2015,Hofmann:2016,Raynaud:2019}. Alternatively, exciton-exciton annihilation (EEA) can be adopted to promote strong photon antibunching \cite{Ma:2015prl,Ishii:2017}. Single-photon emission at room temperature induced by the diffusion-driven EEA among spatially separated free excitons has been demonstrated in undoped air-suspended CNTs \cite{Ishii:2017}.

An additional approach that can possibly control the exciton diffusion and induce strong photon antibunching is the noncovalent functionalization of CNTs by intentional molecular adsorption. Because of the atomically thin nature of CNTs, excitons in air-suspended CNTs are highly sensitive to the dielectric environment, and adsorbed molecules can cause a significant reduction in the excitonic energies \cite{Lefebvre:2004apa,Finnie:2005,Milkie:2005,Ohno:2006prb,Homma:2013,Uda:2018,Machiya:2018,Uda:2018a,Tanaka:2019,Fang:2020}. Meanwhile, the excellent optical properties of the CNTs can be preserved because noncovalent functionalization is less perturbative, and the molecular coverage can be effectively controlled by different adsorption conditions or heating-induced molecular desorption \cite{Finnie:2005,Milkie:2005,Uda:2018,Tanaka:2019}. To induce photon antibunching, a local potential well with a reduced excitonic energy can be created by adsorbing a nanoscale molecular particle onto an air-suspended CNT. Although the potential dip induced by the molecule is expected to be shallow, it should be possible to guide the majority of the free excitons into the potential well and suppress multiphoton emission from spatially separated excitons in the pristine region. As a result of the high density of excitons in the potential well, efficient EEA should occur, and photon antibunching is expected.

By utilizing a type of versatile molecule, pentacene \cite{Dimitrakopoulos:2002,Knipp:2003,Ruiz:2004,Virkar:2010}, we demonstrate directional exciton diffusion and room-temperature single-photon emission in air-suspended CNTs decorated with nanoscale pentacene particles. The pentacene particles are deposited onto individually suspended CNTs by thermal evaporation, where we observe pentacene decoration dependent on the CNT chiral angle. PL imaging and PL excitation (PLE) spectroscopy are performed on such CNTs, and additional peaks with redshifted emission energies are identified after pentacene decoration. Directional exciton diffusion from the undecorated region to the decorated site is demonstrated by PLE spectroscopy on a representative CNT. Excitation-power-dependent PL spectra measured from different emission peaks reveal enhanced EEA at the decorated site, and the exciton trapping rate is calculated to be higher than the detrapping rate at room temperature. Single-photon emission from the redshifted emission peak at the decorated site on the representative CNT is also observed. Additional photon correlation measurements performed on more CNTs indicate that pentacene decoration is on average beneficial for stronger photon antibunching.

\begin{figure}
\includegraphics{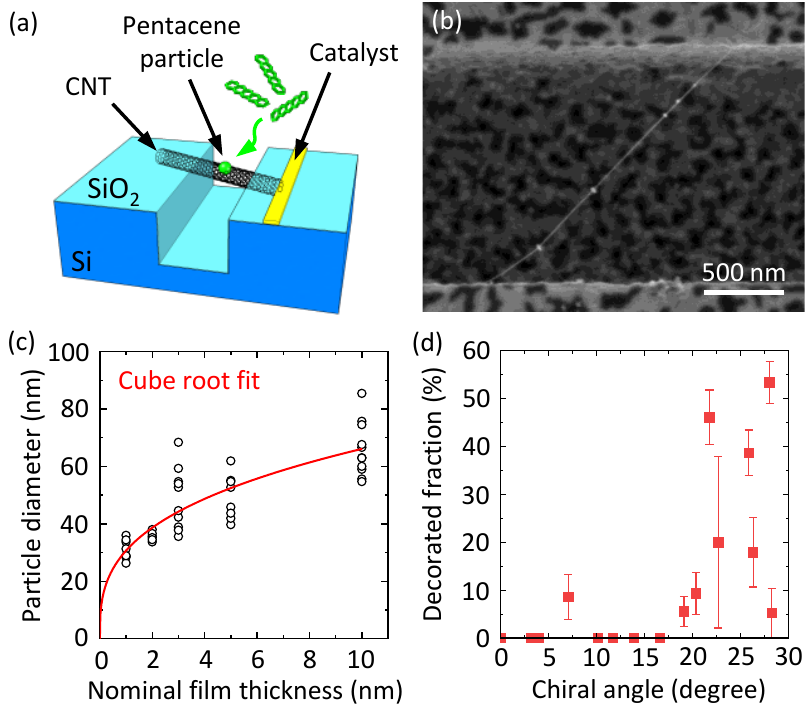}
\caption{\label{Fig1}Pentacene decoration on CNTs and chiral-angle dependence. (a) Schematic showing pentacene adsorption onto an air-suspended CNT on a trenched Si/SiO$_2$ substrate. (b) SEM image of air-suspended CNTs decorated with pentacene particles. (c) Diameters of the decorated pentacene particles as a function of the nominal pentacene film thickness. Red curve is a cube root fit. (d) Percentage of CNTs decorated with pentacene particles with respect to the chiral angles.}
\end{figure}

The air-suspended CNTs [Fig.~\ref{Fig1}(a)] are synthesized on trenched Si/SiO$_2$ substrates \cite{Ishii:2015}. Alignment markers along with trenches with a length of 900~$\mu$m and widths from 0.5 to 4~$\mu$m on the Si substrates are patterned by electron-beam lithography followed by dry etching. A SiO$_2$ film with a thickness of 60--70~nm is then formed upon Si by thermal oxidation. An additional electron-beam lithography process defines the catalyst regions along the edge of the trenches. To synthesize the air-suspended CNTs, catalyst solution of iron acetylacetonate and fumed silica in ethanol is spin-coated onto the catalyst regions on the trenched Si/SiO$_2$ substrates followed by lift-off. CNTs are then grown across the trenches via alcohol chemical vapor deposition at 800$^{\circ}$C for 1~min \cite{Maruyama:2002,Ishii:2015}.

To deposit pentacene onto the air-suspended CNTs, the sample is loaded into a thermal evaporator chamber with a pressure in the order of 10$^{-2}$~Pa. Then pentacene powder (99\%, Sigma-Aldrich) in an alumina crucible is heated by a tungsten filament using a current of 10~A with a typical deposition rate of 0.1--0.5~{\AA}/s. After deposition, amorphous pentacene particles with diameters of tens of nanometers are decorated on the CNTs, as shown in Fig.~\ref{Fig1}(b) and Supplementary Fig.~S1. As the thickness of the deposited pentacene film is monitored by a quartz crystal sensor during the deposition process, the approximate size of the particles can be quantitatively controlled by adjusting the deposition duration. In Fig.~\ref{Fig1}(c), the decorated pentacene particle diameters measured in the scanning-electron microscope (SEM) images are plotted against the nominal thickness of the pentacen film on the substrate. Their relation follows a cube root fit
\begin{equation}
\label{equation1}
d = kt^{1/3},
\end{equation}
where the coefficient $k=30.7$~nm$^{2/3}$, $d$ is the particle diameter, and $t$ is the nominal pentacene film thickness. We note that the actual pentacene film thickness on the substrate is distributed nonuniformly, and there is a discrepancy between the actual film thickness and the nominal thickness (Supplementary Fig.~S2). Nevertheless, the cube root relation is reasonably consistent.

The percentage of CNTs decorated with pentacene particles is obtained by automated PL spectroscopy on about 400 individually suspended CNTs (Supplementary Information). As shown in Fig.~\ref{Fig1}(d), pentacene decoration exhibits a dependency on the CNT species, namely, CNTs with chiral angles larger than 20$^{\circ}$ exhibit a greater probability to be decorated. We speculate that the pentacene decoration is influenced by the orientation of the pentacene molecule relative to that of the CNT. Density-functional theory calculations \cite{Liu:2009} have demonstrated that the ``bridge'' configuration (six-membered rings of pentacene laying over the middle of the C-C bonds in a plane-to-plane configuration) in an armchair CNT (chiral angle 30$^{\circ}$) is energetically favorable for pentacene adsorption, while CNTs with smaller chiral angles are less favorable due to the high curvature along the direction of this configuration.

Before and after pentacene deposition, PL measurements are performed on the same air-suspended CNTs in a home-built confocal microscopy system \cite{Watahiki:2012,Jiang:2015,Ishii:2015,Ishii:2017,Otsuka:2019,Ishii:2019,Fang:2020}. A wavelength-tunable Ti:sapphire laser is used for PL spectroscopy with a continuous-wave (CW) output, while an output of approximately 100~fs pulses and a repetition rate of 76~MHz is used for time-resolved PL measurements. The excitation laser beam with a power $P$ and linear polarization parallel to the CNT axis is focused onto the sample by an objective lens with a numerical aperture of 0.8 (or 0.85 for time-resolved measurements) and a focal length of 1.8~mm. The reflected beam and PL from the sample are collected by the same objective lens and separated by a dichroic mirror. A Si photodiode detects the reflected beam to visualize the trenches on the substrate, and PL is collected by an InGaAs photodiode array attached to a spectrometer. For time-resolved PL and photon-correlation measurements, a fiber-coupled superconducting single-photon detector and a time-correlated single-photon counting module are used. Detection-wavelength-dependent instrument response functions (IRFs) are obtained by supercontinuum white light pulses dispersed by a spectrometer. Photon-correlation measurements are carried out using a Hanbury-Brown-Twiss setup with a 50:50 fiber coupler. Unless otherwise noted, all measurements are carried out at room temperature in dry nitrogen to avoid degradation of pentacene.

\begin{figure}
\includegraphics{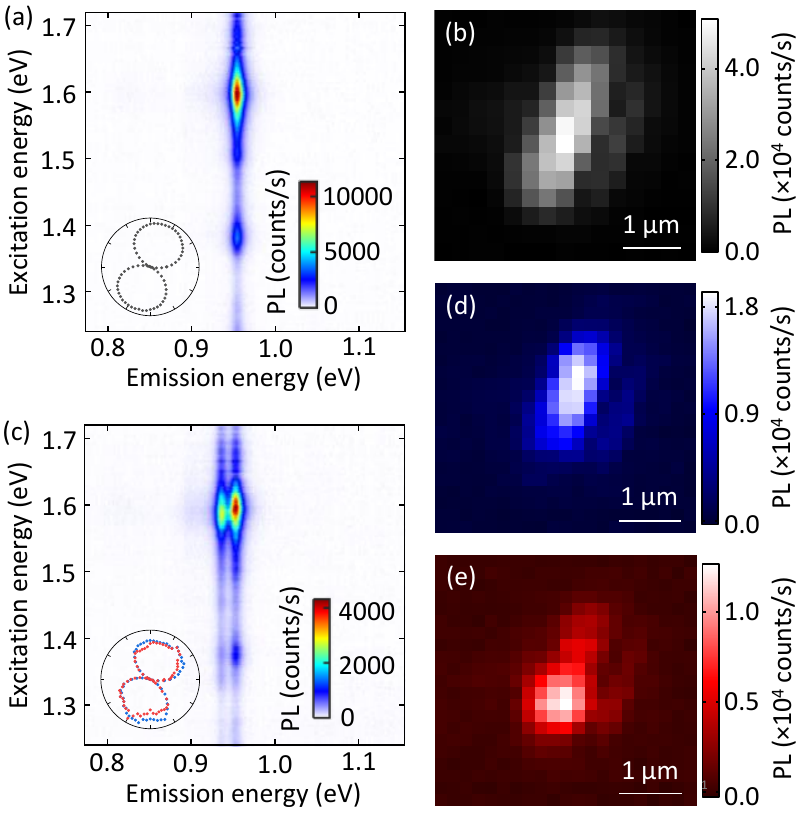}
\caption{\label{Fig2}PL characterization of a typical CNT before and after pentacene decoration. (a) PLE map of a typical (9,7) CNT before pentacene decoration. Inset shows the polarization-dependent integrated PL intensity of the pristine peak (black dots). (b) Integrated PL image of the pristine peak near 0.95~eV. (c) PLE map of the same CNT after pentacene decoration. Inset shows the polarization-dependent integrated PL intensities of the undecorated peak (blue dots) and decorated peak (red dots). Integrated PL images of (d) the undecorated peak near 0.95~eV and (e) decorated peak near 0.94~eV, respectively.}
\end{figure}

After air-suspended CNTs are synthesized on a trenched substrate, automated PL line scans are performed throughout the trenches to locate bright CNTs, then selected CNTs are individually characterized in sequence by polarization-dependent PL, PLE spectroscopy and PL imaging with $P=10$~$\mu$W. A PLE map obtained from a representative pristine CNT (Supplementary Information) is shown in Fig.~\ref{Fig2}(a). This CNT shows bright PL and a peak height of 11680 counts/s, and we name this peak as the ``pristine peak''. The $E_{11}$ and $E_{22}$ resonance energies are at 0.954 and 1.598~eV, respectively. By comparing the resonance energies with tabulated data from air-suspended CNTs \cite{Ishii:2015}, the chirality of this CNT is assigned to (9,7). In Fig.~\ref{Fig2}(b), the PL intensities integrated over a 4~nm spectral window centered at the PL emission energy reveal that the CNT is individual and fully suspended with a length of approximately 3.3~$\mu$m. 

Pentacene with a nominal film thickness of 2.8~nm is then deposited onto this sample after the initial PL characterization. Polarization-dependent PL intensity plots [insets of Fig.~\ref{Fig2}(a) and (c)] are compared before and after pentacene decoration, which allows us to verify that the same CNT has been measured. In the PLE map taken at the pentacene particle, as shown in Fig.~\ref{Fig2}(c), two prominent emission peaks are present. One peak has a peak height of 4416~counts/s, and $E_{11}$ and $E_{22}$ resonance energies at 0.953~eV and 1.596~eV, respectively. Since its resonance energies agree with that of the pristine peak, this peak should correspond to the excitons in the undecorated region on the CNT and is termed as the ``undecorated peak''. The slight redshifts compared to the pristine peak can be caused by fitting errors, while the thin molecular films adsorbed on the undecorated region (Supplementary Fig.~S1) are also speculated to play a role. The additional peak introduced by pentacene decoration has a peak height of 2875~counts/s, and $E_{11}$ and $E_{22}$ resonance energies at 0.938~eV and 1.591~eV, respectively. Both resonance energies are redshifted compared to the pristine peak ($\Delta E_{11}=16$~meV, $\Delta E_{22}=7$~meV), and this peak is named as the ``decorated peak''. Weaker peaks corresponding to the $2u$ state with excitation energies near 1.4~eV are also visible before and after pentacene decoration, which will be discussed later. 

The total peak height of the two prominent peaks is about 62\% of the original peak height in the pristine state, which could suggest efficient EEA and/or PL quenching induced by the adsorbed pentacene molecules. Nevertheless, the reduction in PL intensity  is relatively weak compared to other molecular types that can induce charge transfer \cite{Tanaka:2019}, which is advantageous for exciton diffusion and efficient EEA because long exciton lifetimes and high quantum efficiency can be preserved.

Since an additional emission peak appears after pentacene decoration, separate PL images can be obtained by integrating the PL intensities near emission energies of 0.953~eV and 0.938~eV, respectively [Fig.~\ref{Fig2}(d) and (e)]. Based on the PL images, no significant strain is introduced after pentacene decoration as the shape of the CNT stays the same. The location of the decorated pentacene particle is found to be near the bottom of the CNT as shown in Fig.~\ref{Fig2}(e).

\begin{figure*}
\includegraphics{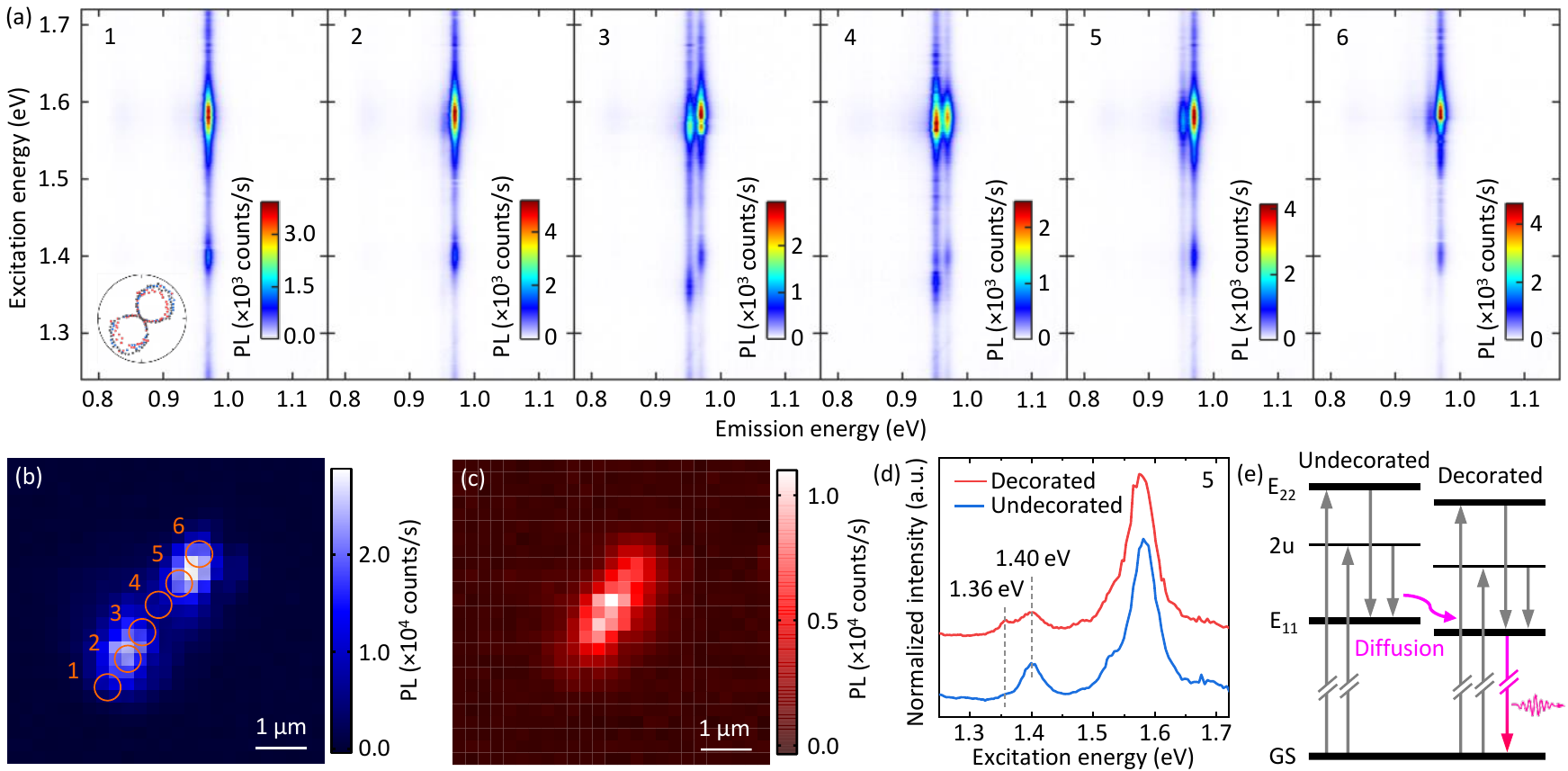}
\caption{\label{Fig3}Location-dependent PLE maps and signatures of exciton diffusion. (a) PLE maps taken at spots 1--6 as marked in (b) along a (9,7) CNT decorated with pentacene. Inset contains the polarization-dependent integrated PL intensities of the pristine peak (black dots), undecorated peak (blue dots) and decorated peak (red dots), respectively. Integrated PL images of the (b) undecorated peak near 0.97~eV and (c) decorated peak near 0.95~eV. (d) PLE spectra from the PLE map taken at spot 5. The red curve corresponds to the decorated peak and the blue curve corresponds to the undecorated peak. (e) Energy level diagram showing the exciton diffusion from the undecorated region to the decorated site.}
\end{figure*}

On another (9,7) CNT decorated with pentacene from the same sample substrate, location-dependent PLE spectroscopy is performed [Fig.~\ref{Fig3}(a)]. The PLE maps are taken from spot 1 to spot 6 along the axis of the CNT as marked by orange circles in the PL image in Fig.~\ref{Fig3}(b). This CNT is decorated with one pentacene particle at the middle as shown in the PL image in Fig.~\ref{Fig3}(c). We note that PL imaging may not be able to resolve multiple closely located pentacene particles since the beam spot size is approximately 1~$\mu$m. In the PLE map taken at spot 1, where the region is not decorated by the pentacene particle, there is a single undecorated peak with $E_{11}$ and $E_{22}$ resonance energies at 0.971~eV and 1.583~eV, respectively. While at spot 4, where the pentacene particle is decorated, two prominent peaks coexist in the PLE map: an undecorated peak with approximately the same energies (0.970~eV and 1.581~eV) as the peak observed at spot 1, and a decorated peak with redshifted $E_{11}$ and $E_{22}$ resonance energies at 0.953~eV and 1.575~eV, respectively. The redshift between the $E_{11}$ resonance energies of the decorated and the undecorated peak at spot 4 is 17~meV, which also corresponds to the depth of the potential well created by pentacene, whereas the redshift between the $E_{22}$ resonance energies is 6~meV. For a CNT under dielectric screening, the amount of redshift in both $E_{11}$ and $E_{22}$ resonance energies should be comparable \cite{Weisman:2003,Fang:2020}. Hence, the relatively small redshift in the $E_{22}$ resonance energies of the pentacene-decorated CNTs cannot be explained by the simple picture of dielectric screening, and the diffusion of excitons should be considered.

The effect of exciton diffusion manifests in the behavior of the weaker $2u$ peaks. As shown in Fig.~\ref{Fig3}(d), the PLE spectra obtained from the PLE map of spot 5 at emission energies corresponding to the undecorated and decorated peaks are depicted as blue and red curves, respectively. Contrary to typical CNTs under dielectric screening \cite{Uda:2018a,Fang:2020}, two $2u$ peaks are present in the decorated PLE spectrum: one $2u$ peak has a redshifted excitation energy at 1.358~eV that is induced by dielectric screening of pentacene; another $2u$ peak has an excitation energy of 1.398~eV that is consistent with the excitation energy 1.401~eV of the $2u$ peak in the undecorated PLE spectrum. The coexistence of the two $2u$ peaks in the decorated PLE spectrum is an indication of directional exciton diffusion, and it can be explained by the energy diagram in Fig.~\ref{Fig3}(e). Excitons at the $2u$ state are generated in the undecorated and decorated regions simultaneously under laser excitation. The excitons from the decorated site are at a lower $2u$ state due to dielectric screening, and they subsequently relax to the lower $E_{11}$ state to recombine radiatively. Meanwhile, the excitons from the undecorated region are at a higher $2u$ state, and they diffuse to the decorated site with a lower $2u$ state followed by radiative recombination from the same lower $E_{11}$ state. This exciton diffusion process gives rise to two $2u$ peaks with the same redshifted $E_{11}$ resonance energy, but with a redshifted and an unshifted excitation energy. Furthermore, because no extra $2u$ peaks are observed in the undecorated PLE spectrum [Fig.~\ref{Fig3}(d)], we conclude that the excitons predominantly diffuse from the undecorated region to the decorated site. Likewise, in the presence of directional exciton diffusion, one should also expect two $E_{22}$ resonance peaks in the decorated PLE spectrum, however, only one redshifted peak is observed as priorly mentioned. It has been reported that the redshift of the $E_{22}$ resonance is two to three times smaller than that of the $2u$ peak in CNTs under dielectric screening \cite{Lefebvre:2008}, and the less redshifted $E_{22}$ resonance of the decorated peak is thus speculated to be the superposition of a $E_{22}$ resonance with a redshifted energy more than 10~meV smaller than 1.583~eV (undecorated peak) due to dielectric screening and an unshifted $E_{22}$ resonance with an energy near 1.583~eV due to directional exciton diffusion.

\begin{figure}
\includegraphics{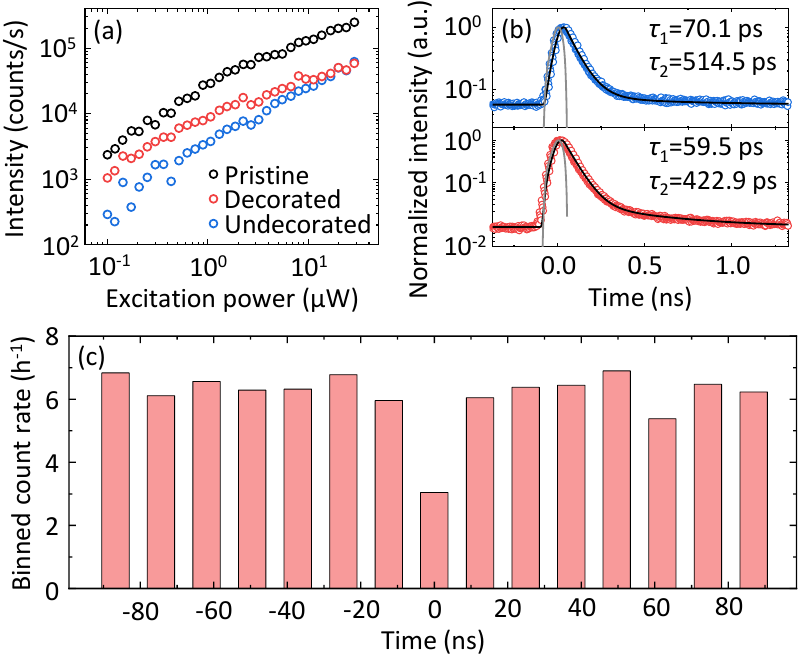}
\caption{\label{Fig4}Power-dependent and time-resolved PL properties. (a) Excitation-power dependence of the PL intensities corresponding to the pristine peak (black circles), decorated peak (red circles) and undecorated peak (blue circles). The excitation energy is 1.59~eV. (b) PL decay curves taken at the pentacene-decorated site on the CNT using a bandpass filter. Top panel: undecorated peak at 0.97~eV. Bottom panel: decorated peak at 0.95~eV. An excitation energy of 1.57~eV and pulsed $P=50$~nW are used. (c) Binned count-rate histogram taken with an excitation energy of 1.57~eV and pulsed $P=10$~nW at the pentacene-decorated site using a bandpass filter. The data are accumulated from measurements performed for 33~h in an ambient condition.}
\end{figure}

To understand the exciton dynamics in the pentacene-decorated CNTs, we perform excitation-power-dependent PL measurements at the particle [spot 4 in Fig.~\ref{Fig3}(b)] on the same CNT where the directional exciton diffusion has been demonstrated. PL intensities as a function of the excitation power for each emission peak before and after pentacene decoration are shown in Fig.~\ref{Fig4}(a). The curves corresponding to the pristine peak and the undecorated peak show a similar behavior throughout the power range, but the curve for the decorated peak deviates to a lower slope above 0.7~$\mu$W. It is known that the PL intensity and the excitation power follow a linear relationship at low powers, and the relationship becomes sublinear at high powers due to EEA \cite{Ishii:2015}. Hence, the early deviation of the decorated peak intensity at lower powers is a signature of more efficient EEA owning to higher density of excitons at the decorated site than in the undecorated region.

PL decay curves corresponding to the undecorated and decorated emission peaks are also measured at spot 4 by using a bandpass filter (Supplementary Fig.~S3), and the exciton lifetimes are extracted from the PL decay curves by biexponential fitting convoluted with the IRF [Fig.~\ref{Fig4}(b)]. Two decay components are obtained in each PL decay curve: a fast component with a lifetime $\tau_1$ associated with bright excitonic states that contribute significantly to the emission intensity, and a slow component with a lifetime $\tau_2$ associated with dark excitonic states \cite{Mortimer:2007,Berciaud:2008,Ishii:2019}. The difference between $\tau_1$ of the undecorated and decorated peaks is not clearly distinguishable, which could be due to the fact that exciton trapping and detrapping occur faster than exciton recombination \cite{Sykes:2019}. Fast exciton trapping and detrapping are also evidenced by the relatively bright decorated peak in the PLE map [Fig.~\ref{Fig3}(a)] considering the small diameter of the pentacene particle ($<$100~nm) with respect to the 1~$\mu$m laser spot size. Supposedly, in a scenario where excitons recombine earlier than they engage in the trapping and detrapping process, the decorated peak intensity should become much weaker since only a small number of excitons are excited at the nanoscale decorated site.

Finally, photon-correlation measurements are conducted at the decorated site on this CNT at ambient condition with the decorated emission peak isolated by the bandpass filter. The autocorrelation histogram is evaluated by subtracting the dark counts and binning each peak, as shown in the binned count-rate histogram in Fig.~\ref{Fig4}(c). The zero-delay second-order intensity correlation $g^{(2)}(0)$ is obtained by calculating the ratio between the peak count at zero time delay and the average count of the other peaks. For this pentacene-decorated CNT, $g^{(2)}(0)$ is 0.48 at $P=10$~nW, indicating single-photon emission at room temperature.

Quantitatively, the strength of exciton trapping by the pentacene particle can be evaluated from the ratio between the trapping rate and the detrapping rate. Assuming a single potential well is created by the pentacene particle on this CNT, and excitons can occupy either the decorated state or the undecorated state, the trapping rate to detrapping rate ratio is then determined by the Boltzmann factor as
\begin{equation}
\label{equation2}
\frac{k_{\text{trap}}}{k_{\text{detrap}}}=\exp(\frac{\Delta E}{k_{\text{B}}T}),
\end{equation}
where $\Delta E$ is the energy difference between the undecorated state and the decorated state, $k_{\text{B}}$ is the Boltzmann constant, and $T$ is the temperature. Using $\Delta E=17$~meV as obtained from the PLE map in Fig.~\ref{Fig3}(a), the ratio between the trapping rate and the detrapping rate at room temperature is 1.9. This suggests that the potential well created by pentacene can induce exciton trapping at room temperature. 

Furthermore, the validity of our calculation can be verified by using the steady-state rate equation for excitons within the laser spot
\begin{equation}
\label{equation3}
g+k_{\text{detrap}}N^*-\frac{N}{\tau}-k_{\text{trap}}N=0,
\end{equation}
\begin{equation}
\label{equation4}
g^*+k_{\text{trap}}N-\frac{N^*}{\tau^*}-k_{\text{detrap}}N^*=0,
\end{equation}
where $g$ ($g^*$) is the generation rate of excitons in the undecorated (decorated) state, $N$ ($N^*$) is the exciton population in the undecorated (decorated) state, and $\tau$ ($\tau^*$) is the lifetime of excitons in the undecorated (decorated) state. Under CW excitation, the generation rate and the recombination rate are considered equal, and the trapping and detrapping rates are sufficiently faster than exciton recombination as we have found from the PL decay curves. Hence, the terms containing $g$ ($g^*$) and $\tau$ ($\tau^*$) can be considered negligible, and the ratio between $N^*$ and $N$ is simply expressed as
\begin{equation}
\label{equation5}
\frac{N^*}{N}=\frac{k_{\text{trap}}}{k_{\text{detrap}}}.
\end{equation}
Experimentally, the exciton population ratio can be estimated by comparing the emission peak areas of the decorated peak and the undecorated peak. When we consider the PLE map taken at spot 4 in Fig.~\ref{Fig3}(a), the peak area ratio $I^*⁄I=N^*⁄N\approx1.4$. This value obtained from the emission peak area ratio is in reasonable agreement with the exciton population ratio 1.9 derived from the exciton trapping rate and the detrapping rate in Eq.~(\ref{equation2}), indicating that our trapping rate calculation is valid. The difference between the two values is likely caused by a lower emission quantum efficiency of the decorated peak which is not considered in this model. By comparing the two values, the decrease in the emission quantum efficiency due to pentacene decoration is estimated to be 26\% in this CNT. Alternatively, discrepancy may also occur if the actual trapping and detrapping rates are not fast enough for $g$ ($g^*$) and $1/\tau$ ($1/\tau^*$) to be negligible.

Additional time-resolved PL and photon-correlation measurements are carried out on 8 more pentacene-decorated CNTs, where separate sets of data are acquired from the undecorated peak and the decorated peak (Supplementary Table~S1). No clear correlation can be found between the exciton lifetimes calculated from the undecorated and decorated peaks, again suggesting that the trapping and detrapping process is faster than exciton recombination. Meanwhile, values of $g^{(2)}(0)$ obtained at pulsed $P=20$~nW from the undecorated peak and decorated peak are also compared. Various factors are known to have a potentially negative impact on the results, such as PL blinking, pentacene degradation over time, shortened exciton diffusion length limited by the thin molecular films in the undecorated region, as well as defocusing and misalignment due to occasional sample drifting. Nevertheless, clear antibunching is observed from both peaks, and 6 out of 8 CNTs exhibit a lower $g^{(2)}(0)$ from the decorated peak with a minimum value reaching 0.38. The average $g^{(2)}(0)$ from the decorated peak is also noticeably lower than that obtained from pristine air-suspended CNTs under a similar condition \cite{Ishii:2017}. Overall, it suggests that the decorated pentacene particles are generally beneficial for stronger photon antibunching in air-suspended CNTs.

In conclusion, we have demonstrated directional exciton diffusion and room-temperature single-photon emission from air-suspended CNTs decorated with pentacene particles. Typical decorated pentacene particles have a diameter less than 100~nm, and the particle size can be controlled by adjusting the deposition parameters during thermal evaporation. Pentacene decoration is found to show a dependency on the CNT species, where CNTs with chiral angles larger than 20$^{\circ}$ are more likely to be decorated. CNTs are characterized by PL spectroscopy, and the typical depth of the potential well created by pentacene decoration is estimated to be less than 20~meV. The location of the decorated particle can be visualized in the PL image, and directional exciton diffusion from the undecorated region to the decorated site is demonstrated in the PLE map where an additional $2u$ peak with redshifted $E_{11}$ resonance emerges. Enhanced EEA induced by directional exciton diffusion to the decorated site is confirmed in the representative CNT by excitation-power-dependent PL spectra, and the exciton trapping rate at the decorated site is estimated to be higher than the detrapping rate at room temperature. From photon-correlation measurements, single-photon emission from excitons in the decorated site is verified. More CNTs are also found to show enhance photon antibunching induced by pentacene decoration. To further modify the energy landscape, local laser heating can be adopted to pattern the distribution of the decorated particles. Moreover, a molecular type that can introduce deeper traps and cryogenic temperatures can be considered to give rise to an even lower $g^{(2)}(0)$. Our results indicate that noncovalent functionalization of CNTs using nanoscale molecular particles is a promising top-down approach for energy landscape modification and quantum emission at room temperature.

\begin{acknowledgments}
This work is supported in part by MIC (SCOPE 191503001), JSPS (KAKENHI JP20H02558, JP19J00894, JP20J00817, JP20K15121, JP20K15137, JP20K15112 and JP20K15199), and MEXT (Nanotechnology Platform JPMXP09F19UT0079). K.~O. and D.~Y. are supported by the JSPS Research Fellowship. D.~K. is supported by the RIKEN Special Postdoctoral Researcher Program. We acknowledge RIKEN Materials Characterization Team for access to the transmission electron microscope, and RIKEN Advanced Manufacturing Support Team for technical assistance.
\end{acknowledgments}

\end{document}